\documentclass[12pt]{iopart}
\usepackage[utf8]{inputenc}
\expandafter\let\csname equation*\endcsname\relax
\expandafter\let\csname endequation*\endcsname\relax
\usepackage{amssymb,amsmath}
\usepackage{graphicx}
\usepackage{subcaption}
\usepackage{natbib}
\usepackage[colorlinks=true,linkcolor=black,citecolor=black,urlcolor=blue,pdfborder={0 0 0}]{hyperref}

\usepackage{xcolor}
\newcommand{\cb}{}

\begin{document}

\title[]{Modeling the propagation of tumor fronts with shortest path and diffusion models -- implications for the definition of the clinical target volume}

\author{Thomas Bortfeld$^1$ {\cb and Gregory Buti$^{2,1}$}}

\address{$^1$Massachusetts General Hospital and Harvard Medical School, Department of Radiation Oncology, Division of Radiation Biophysics, 100 Blossom St, Boston, MA 02114, United States of America

{\cb $^2$UCLouvain, Institute of Experimental and Clinical Research, Center of Molecular Imaging, Radiotherapy and Oncology, Avenue Hippocrate 54 - box B1.54.07, 1200 Brussels, Belgium}
}

\ead{tbortfeld@mgh.harvard.edu}
\vspace{10pt}
\begin{indented}
\item[]\today
\end{indented}

\begin{abstract}

{\em Objective}: The overarching objective is to make the definition of the clinical target volume (CTV) in radiation oncology less subjective and more scientifically based. The specific objective of this study is to investigate similarities and differences between two methods that model tumor spread beyond the visible gross tumor volume (GTV): 1. The shortest path model, which is the standard method of adding a geometric GTV-CTV margin, and 2. The reaction-diffusion model. 

{\em Approach}: These two models to capture the invisible tumor ``fire front’’ are defined and compared in mathematical terms. The models are applied to geometric example cases that represent tumor spread in non-uniform and anisotropic media with anatomical barriers. 

{\em Main Results}: The two seemingly disparate models bring forth traveling waves that can be associated with the front of tumor growth outward from the GTV. The shape of the fronts is similar for both models. Differences are seen in cases where the diffusive flow is reduced due to anatomical barriers, and in complex spatially non-uniform cases. The diffusion model generally leads to smoother fronts. The smoothness can be controlled with a parameter defined by the ratio of the diffusion coefficient and the proliferation rate. 

{\em Significance}: 
Defining the CTV has been described as the weakest link of the radiotherapy chain. There are many similarities in the mathematical description and the behavior of the common geometric GTV-CTV expansion method, and the definition of the CTV tumor front via the reaction-diffusion model. Its mechanistic basis and the controllable smoothness make the diffusion model an attractive alternative to the standard GTV-CTV margin model.
\end{abstract}

\vspace{2pc}



\section{Introduction}
The Clinical Target Volume (CTV) is a tissue volume that contains a demonstrable Gross Tumor Volume (GTV) and/or subclinical malignant disease that must be eliminated \citep{ICRU62}. Prescribing, recording and reporting photon beam therapy beyond the visible GTV to include the CTV and the planning target volume was first formalized by the ICRU (International Commission on Radiation Units and measurements) in their Report 50 \citep{ICRU50} and further refined in Report 62 \citep{ICRU62}. Subsequent ICRU reports applied these concepts to the treatment with electrons (Report 71), protons (Report 78), intensity-modulated radiation therapy (Report 83), brachytherapy (Report 89), stereotactic treatments (Report 91), and light ion therapy (Report 93). Delineating the tumor target volume and in particular the CTV was found to be the weakest link in the radiotherapy chain \citep{njeh2008}. It has recently been identified as one of the four grand challenges for medical physics in radiation oncology \citep{fiorino2020}. The overall mission is to complement the art of defining the CTV with a geometrical or physical model-based approach. 

The infiltrative growth of  tumors into surrounding healthy tissues, especially in the case of aggressive tumors such as  glioblastoma multiforme (GBM), has been compared to the moving front of a wildfire \citep{woodward1996}. One of the big challenges in radiation treatment planning is that the fire front is invisible. Defining the CTV means estimating the position of an invisible fire front. The art of defining the CTV has been described in textbooks \citep{grosu2015}. For specific disease sites, consensus guidelines have been developed \citep{niyazi2016, gregoire2018, wang2011, salerno2021}. They normally recommend expanding the GTV by a fixed margin, while avoiding anatomical barriers that serve as impenetrable walls for tumor cell spread \citep{shusharina2020}. In the case of directional tissues such as white matter tracts in the brain, or muscle fibers, the GTV-CTV expansion guidelines may recommend larger expansions in the direction of preferred spread \citep{jordan2019, wang2011, salerno2021}. 

Parallel to the development of the geometric GTV-CTV expansion methods, more mechanistic approaches to model the propagation of the tumor front into normal tissues have been developed. Among the most well-known approaches of this type are diffusion models where the movement of {\cb tumor cells} depends on the  gradients of the tumor cell density. These models were advanced in particular in glioma \citep{tracqui1995, woodward1996, burgess1997, murray2001, swanson2003,  konukoglu2009, konukoglu2010, unkelbach2014, lujan2016}. While  diffusion models yield the tumor cell density at any point in space and time, the tumor front can be derived as an iso-surface (level set) of the 3D density map. The front propagates outward from the GTV through the surrounding tissues as a function of time.   
This allows for a direct comparison of these iso-cell-density fronts with the iso-distance fronts obtained from the geometric expansion method above. 

The purpose of this {\cb paper} is to do such a comparison of these two seemingly disparate approaches to model tumor front propagation: geometric expansion and diffusion. Others have addressed this question and overlaid fronts (contours) resulting from the two methods \citep{konukoglu2009}. {\cb In a recent publication, \cite{hager2022} compared cell density iso-lines resulting from a diffusion model with manually defined CTVs using standard geometric expansion. However, }a systematic comparison rooted in the governing equations appears to be missing.

\section{Methods}
\subsection{Geometric expansion from GTV to CTV}
Let $S(\boldsymbol{r})$ measure the shortest distance of any point $\boldsymbol{r}= (x, y, z)$  from the GTV surface. Then, for any distance $\sigma$ outside the GTV boundary, the CTV surface at distance $\sigma$ (that is, $\text{CTV}(\sigma)$) is defined as the iso-distance level set $S(\boldsymbol{r}) = \sigma$, i.e., the set of all points $\boldsymbol{r}: S(\boldsymbol{r}) = \sigma$.

The iso-distance front propagates outward along the gradient $\boldsymbol{\nabla} S(\boldsymbol{r})$ of the distance function such that \begin{equation}
    \| \boldsymbol{\nabla} S(\boldsymbol{r}) \|^2 = g(\boldsymbol{r}), \label{EQ_Eik}
\end{equation} 
where $g(\boldsymbol{r})$ is the ``resistance" to tumor spread at point $\boldsymbol{r}$ and $\| \cdot \|$ is the Euclidean norm. {\cb In standard treatment planning, the value of $g(\boldsymbol{r})$ is generally assumed to be $g(\boldsymbol{r})=1$, resulting in a uniform expansion of the GTV. In obstacles/barriers to tumor spread, $g(\boldsymbol{r}) = \infty$.  If certain tissues are known or assumed to be more or less resistant, one can set the values to  $g(\boldsymbol{r})>1$ or $g(\boldsymbol{r})<1$, see also \cite{shusharina2020}.}
Equation (\ref{EQ_Eik}) is an Eikonal equation, which lends itself to a solution via Dijkstra-like ordered upwind algorithms \citep{tsitsiklis1995}, specifically with the so-called fast marching method (FMM) \citep{sethian1996, sethian2003}. 

In the most general anisotropic case where the resistance to tumor spread depends on the direction of the spread, {\cb the distance increment (arc segment) $ds$ depends on the infinitesimal position increment $d\boldsymbol{r}'=(dx,dy,dz)$ as 
\begin{equation}
(ds)^2 = d\boldsymbol{r}' \cdot  \mathcal{G}(\boldsymbol{r}) \cdot d\boldsymbol{r}. \label{EQ_aniso}
\end{equation}
Here $\mathcal{G}(\boldsymbol{r})$ is a metric tensor, a positive definite matrix.} Throughout this paper the prime indicates the transposed vector, i.e., $d\boldsymbol{r}'$ is the corresponding row vector to the column vector $d\boldsymbol{r}$. {\cb In the appendix it is shown that the anisotropic generalization of the Eikonal equation (\ref{EQ_Eik}) is
\begin{equation}
\boldsymbol{\nabla}' S(\boldsymbol{r}) \cdot  \mathcal{G}^{-1}(\boldsymbol{r}) \cdot \boldsymbol{\nabla} S(\boldsymbol{r}) = 1, \label{EQ_Riemann_comp}
\end{equation}
where $\mathcal{G}^{-1}(\boldsymbol{r})$ is the inverse of the metric tensor $\mathcal{G}(\boldsymbol{r})$ above. {\cb This equation needs to be solved for $S(\boldsymbol r)$. It} can be written in a form that more closely resembles equation (\ref{EQ_Eik}):
\begin{equation}
    \| \boldsymbol{\nabla} S(\boldsymbol r)\| 
    = \frac{1}{\sqrt{\boldsymbol{n}' (\boldsymbol{r}) \cdot \mathcal{G}^{-1}(\boldsymbol{r}) \cdot \boldsymbol{n} (\boldsymbol{r})
    }}. \label{EQ_Riemann_Eik}
\end{equation}
Here} $\boldsymbol{n} (\boldsymbol{r})$ is the unit vector in the direction of the propagation of the iso-distance front at point $\boldsymbol{r}$: $\boldsymbol{n}(\boldsymbol{r}) = \boldsymbol{\nabla} S(\boldsymbol{r}) / \| \boldsymbol{\nabla} S(\boldsymbol{r}) \|$.

\subsubsection{Numerical solution.}
The Eikonal equation (\ref{EQ_Riemann_Eik}) is solved {\cb for $S(\boldsymbol r)$} using the Hamiltonian Fast Marching library in Python   \citep{mirebeau2018}. {\cb Our metric tensor $\mathcal{G}$ is the Riemann metric in the formalism of \cite{mirebeau2018}. 

The Fast Marching Method (FMM) starts from the known solution $S(\boldsymbol r)=0$ on the surface of the GTV. The algorithm then proceeds step by step through the neighboring layers of voxels outside of the GTV, and ultimately calculates the distance from the GTV for every voxel in the entire volume. FMM is a one-pass algorithm, which makes it as fast as the Dijkstra algorithm, without suffering from its large discretization errors.} 

\subsection{Diffusion models}
A common model describing the growth and spread of such tumors is the reaction-diffusion model \citep{tracqui1995,swanson2003,konukoglu2009,unkelbach2014}:

\begin{equation}
  \frac{\partial}{\partial t} u(\boldsymbol{r}, t) = 
  \boldsymbol{\nabla}' \cdot \mathcal{D}(\boldsymbol{r}) \cdot \boldsymbol{\nabla} u(\boldsymbol{r},t) 
  + \rho u(\boldsymbol{r},t) \left(1-\frac{u(\boldsymbol{r},t)}{u_{\max}} \right),
  \label{EQ_fisher}
\end{equation}
where $u(\boldsymbol{r},t)$ is the tumor cell density at location $\boldsymbol r$ and time $t$, $\mathcal{D}(\boldsymbol{r})$ is the diffusion tensor, {\cb $\rho \ge 0$} the proliferation rate, and $u_{\max}$ is the {\cb finite} carrying capacity.  The first term is the spread to neighboring locations, and the second term is the increase in cell density at a location where the tumor is already present.

The asymptotic behavior of this Fisher-Kolmogorov equation is given by a traveling wave solution: Over time, the tumor front moves outwards {\cb from the GTV} into the surrounding healthy tissue at a velocity $v(\boldsymbol{r})$, which relates to the model parameters via \citep{konukoglu2009,unkelbach2014} 
\begin{equation}
    v(\boldsymbol{r}) \approx 2 \sqrt{\boldsymbol{n}'(\boldsymbol{r}) \cdot \mathcal{D}(\boldsymbol{r}) \cdot \boldsymbol{n}(\boldsymbol{r}) \, \rho}, \label{EQ_speed}
\end{equation}
where $\boldsymbol n (\boldsymbol{r})$ is again the unit vector normal to the front at point $\boldsymbol{r}$, pointing in the direction of the motion of the front. 

Under the assumption of isotropic diffusion where $\mathcal{D}$ is a diagonal matrix with diagonal elements equal to the diffusion coefficient $d$, the propagation of the wave front is governed by (i) the product and (ii) the ratio of the diffusion coefficient and the proliferation rate:
\begin{enumerate}
\item The parameter $v = 2 \sqrt{ d \rho }$ is a measure of the velocity of the wave propagation as above.
\item The parameter $q = \sqrt{d / \rho}$ is a measure of the ``infiltration length'', over which the front drops from  the carrying capacity $u_{\max}$ to 0. This parameter also determines the minimum radius of curvature of the front (i.e., the smoothness), as we will see.
\end{enumerate}
For example, in GBM, with an infiltration length of $q = 5$ mm in white matter \citep{unkelbach2014} and a proliferation rate in the order of $\rho =1\%$/day, the velocity is $v = 0.1$~mm/day (3 mm per month) and the diffusion coefficient is $d = 0.25\, \text{mm}^2$/day. 

The emergence of the traveling wave solution can be motivated by observing that in the quasi one-dimensional case where there is no variation in $y$ and $z$ direction, the Ansatz $u(\boldsymbol{r},t) = u(\xi)  = u(x-vt)$  solves equation (\ref{EQ_fisher}) with $u(\xi) = \exp{\left(-\xi/q \right)}$ for large $\xi$, thus describing a traveling planar wave. Note that a better approximation may be obtained by modeling the velocity as time-dependent \citep{ebert2000, konukoglu2009}.

Now {\cb we will come} back to the general spatially non-uniform and anisotropic case. Let $T(\boldsymbol{r})$ be the arrival time at which the tumor front reaches point $\boldsymbol{r}$.
The tumor wave front at time $\tau$ is the level set $T(\boldsymbol{r})=\tau$. Locally at point $\boldsymbol{r}$ the gradient $\boldsymbol{\nabla} T (\boldsymbol{r})$ is perpendicular to the wave front. The unit normal vector is thus $\boldsymbol n(\boldsymbol{r}) = \boldsymbol{\nabla} T (\boldsymbol{r}) / \| \boldsymbol{\nabla} T(\boldsymbol{r}) \|$. This yields the  Eikonal equation: 
\begin{equation}
    \| \boldsymbol{\nabla} T(\boldsymbol r)\| = \frac{1}{v(\boldsymbol r)} 
    \approx \frac{1}{2 \sqrt{\boldsymbol{n}'(\boldsymbol{r}) \cdot \mathcal{D}(\boldsymbol{r}) \cdot \boldsymbol{n}(\boldsymbol{r})  \, \rho}}{\cb ,}
    \label{EQ_Diff_Eik}
\end{equation}
or
\begin{equation}
    \boldsymbol{\nabla}' T(\boldsymbol{r}) \cdot \mathcal{D}(\boldsymbol{r}) \cdot \boldsymbol{\nabla} T(\boldsymbol{r})  \approx \frac{1}{4 \rho}.
    \label{EQ_Diff_comp}
\end{equation}

Comparing the two approximate relationships (\ref{EQ_Diff_Eik}) and (\ref{EQ_Diff_comp}) above with equations (\ref{EQ_Riemann_Eik}) and 
(\ref{EQ_Riemann_comp}), we see that the diffusion tensor $\mathcal{D}(\boldsymbol{r})$ corresponds with the (inverse) metric tensor $\mathcal{G}^{-1}(\boldsymbol{r})$, save for a constant factor.

\subsubsection{Numerical solution.}
Numerically we evolve the tumor cell density over time, starting from the following initial distribution where $u$ is at the carrying capacity $u_{\max}$ within the GTV, and $0$ elsewhere:
\begin{equation}
    u(\boldsymbol{r}, t_0)=\begin{cases}
    u_{\max}&  \text{if $\boldsymbol{r} \in $ GTV},\\
    0 &  \text{otherwise}.
\end{cases}
\end{equation}
{\cb The time is fully discretized at $t_n$ with a constant time interval $\Delta t = t_{n+1} - t_n$.}
The cell densities at later times $t_{n+1}$ are obtained by numerically integrating the diffusion equation using the forward time centered space (FTCS) scheme \citep{press2007}: 
\begin{equation}
  u(\boldsymbol{r}, t_{n+1}) = u(\boldsymbol{r}, t_{n}) + \Delta t \left(
  \boldsymbol{\nabla}' \cdot \mathcal{D}(\boldsymbol{r}) \cdot \boldsymbol{\nabla} u(\boldsymbol{r},t_n) 
  + \rho u(\boldsymbol{r},t_n) \left(1-\frac{u(\boldsymbol{r},t_n)}{u_{\max}} \right) \right).
\end{equation}
The space is fully discretized on an equi-spaced grid $\boldsymbol{r} = (x_i, y_j, z_k)$ with spacing $(\Delta x, \Delta y, \Delta z)$. The gradient is approximated by finite differences. For example, in the first component we have 
\begin{equation}
\boldsymbol{\nabla}_i \, u(x_i,y_j,z_k,t_n) = \frac{u(x_{i+1}, y_j, z_k, t_n) - u(x_{i-1}, y_j, z_k, t_n)}{2\Delta x}.
\end{equation}

This FTCS model is implemented in Python. 
Note that in order to avoid oscillations, the time step $\Delta t$ has to satisfy the following Courant condition, {\cb depending on the grid spacing and the diffusion coefficient $d$:}
\begin{equation}
    \Delta t < \frac{(\min(\Delta x, \Delta y, \Delta z))^2}{\max(d)}.
    \label{EQ_timestep}
\end{equation}
Furthermore, the grid spacing in $(\Delta x, \Delta y, \Delta z)$ has to be smaller than $q$.

\section{Results}
\begin{figure}[ht]
\centering

\begin{subfigure}{0.6\textwidth}
\centering
\includegraphics[width=\textwidth]{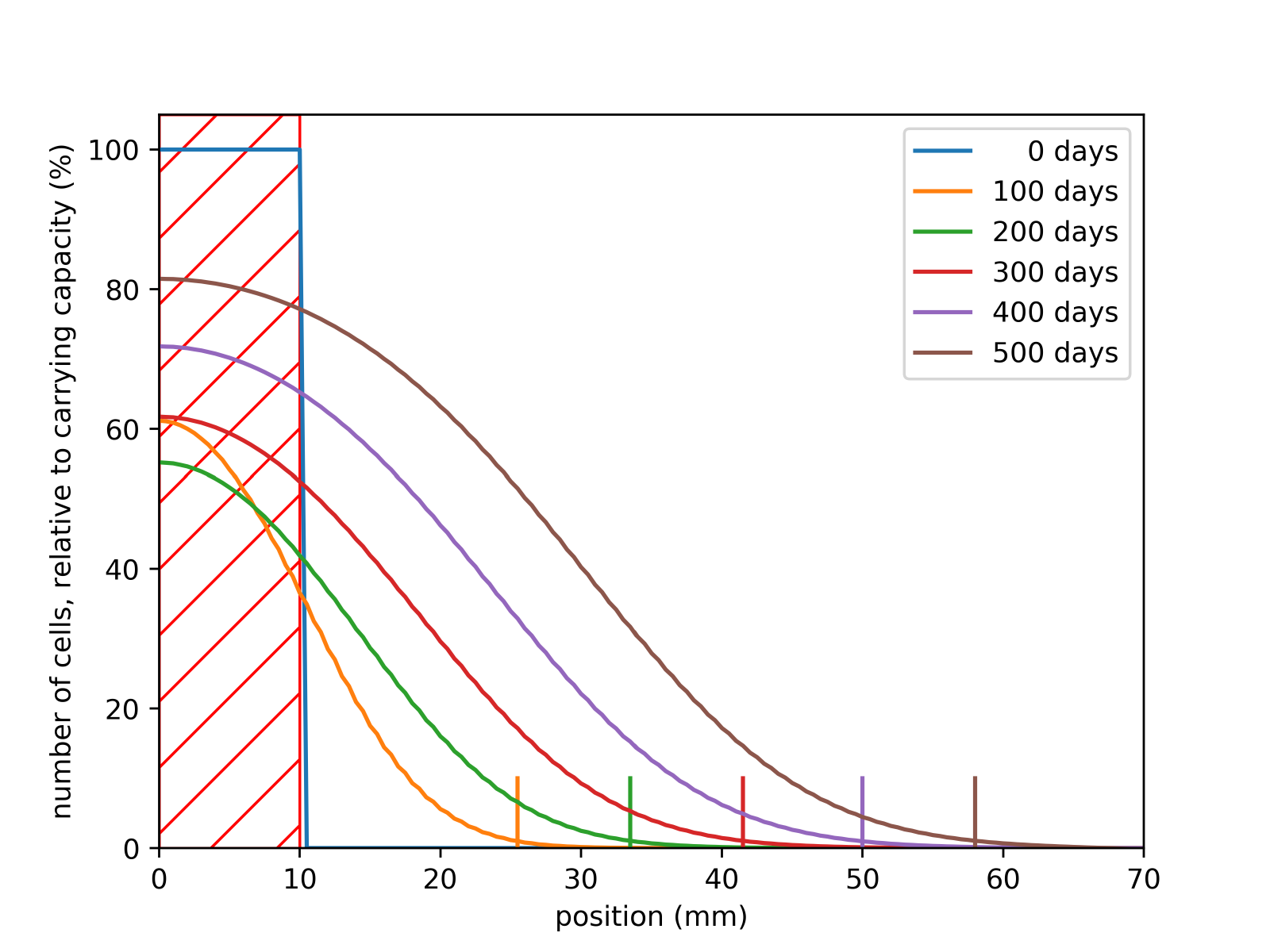}
\caption{}
\end{subfigure}

\begin{subfigure}{0.6\textwidth}
\centering
\includegraphics[width=\textwidth]{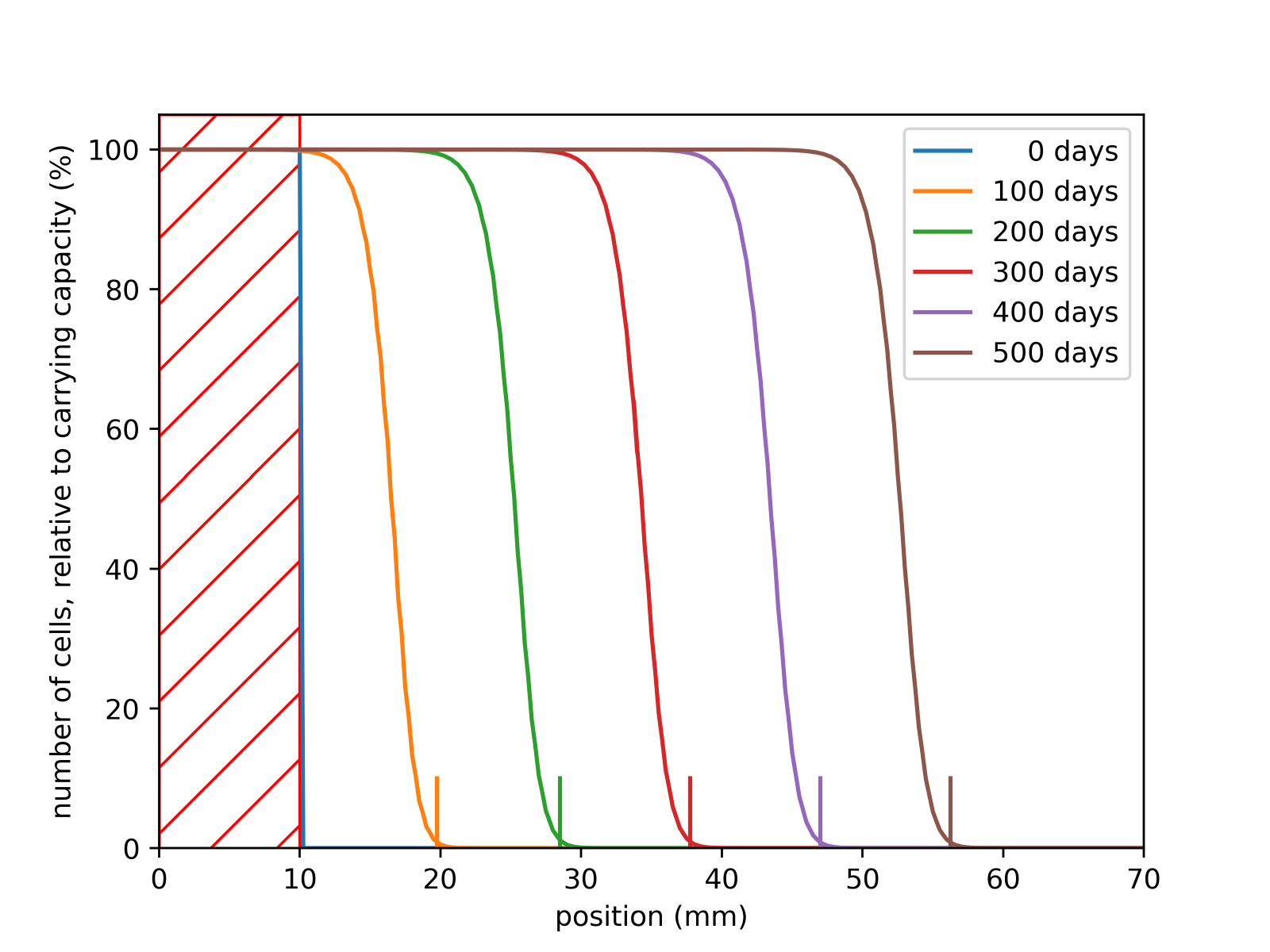}
\caption{}
\end{subfigure}

\caption{\cb Diffusion of tumor cells in a spherically symmetric case, starting at $t=0$ from a spherical GTV with a radius of 10 mm (red diagonal lines). The short vertical lines represent the fronts at 1\% cell density at the different time points. The $q$ parameter, a measure of the ``infiltration length'' of the fronts, was set to $q=5$~mm in (a) and to $q=0.5$~mm in (b), resulting in steeper fronts. \label{FIG_1}}
\end{figure}

{\cb Let us first apply the diffusion model to a spherically symmetric case where the GTV is a sphere with a radius of 10 mm.} The initial condition is a cell density in the GTV at 100\% of $u_{\max}$, and zero everywhere else, shown by the blue curve in figure \ref{FIG_1}. In this example, as in all subsequent examples, the reference diffusion coefficient is set to $d = 0.25\, \text{mm}^2/\text{day}$ and the proliferation rate to 1\%/day. These values are informed by the diffusion of GBM, but they are chosen here primarily for illustrative purposes: Based on equation (\ref{EQ_speed}) they lead to an expected speed of the wave front of {\cb approximately} $v=0.1$~mm/day, which is 1~cm in 100 days. The plot confirms this. In figure \ref{FIG_1}(b) the infiltration length parameter $q=\sqrt{d / \rho}$ is decreased from 5 mm to 0.5 mm by a ten-fold decrease of the diffusion coefficient to $d = 0.025\, \text{mm}^2/\text{day}$ and an (unrealistic, just for illustration purposes) ten-fold increase of the proliferation rate to 10\%/day. 

\begin{figure}[ht]
\centering
\includegraphics[width=0.7\textwidth]{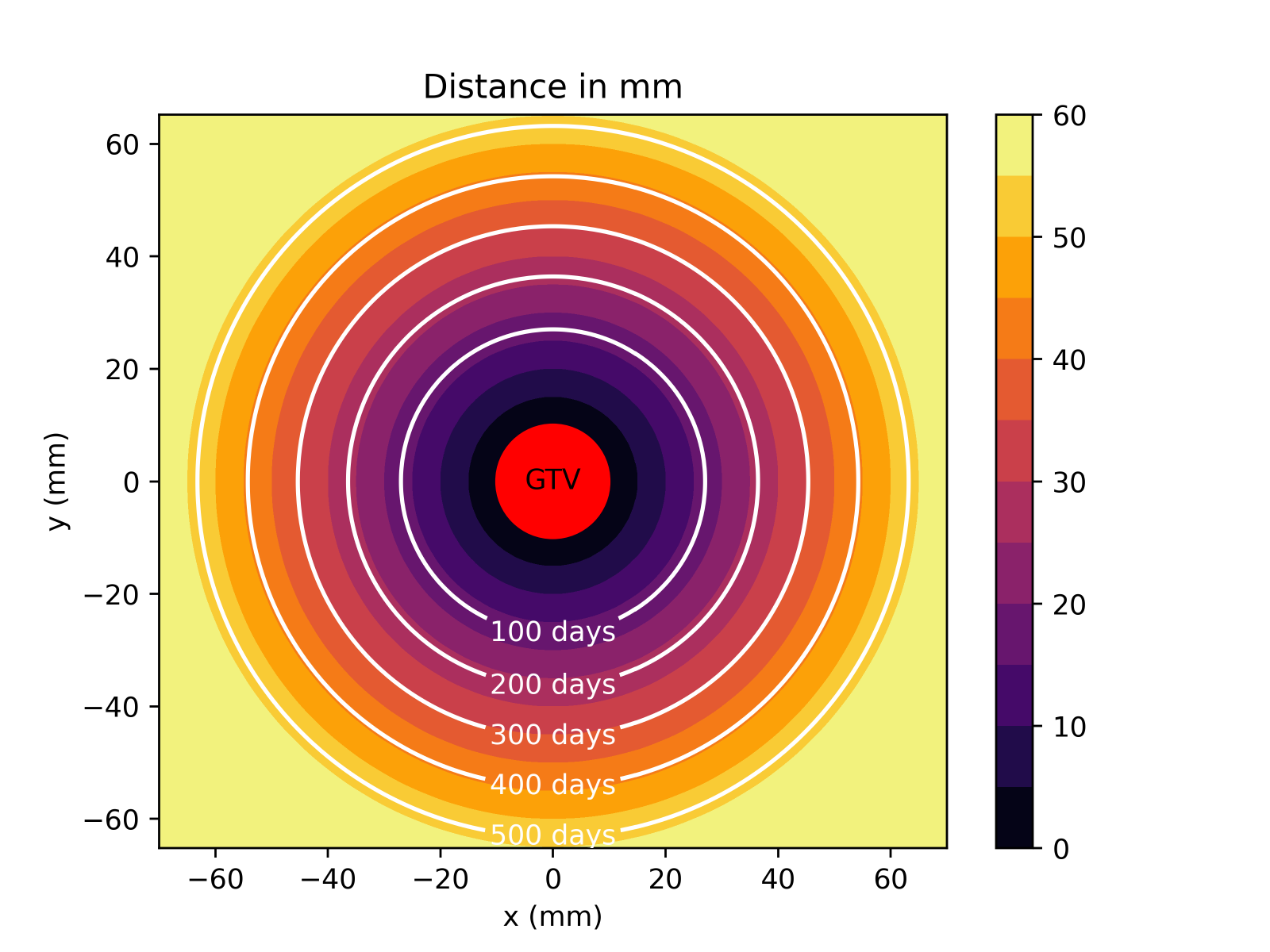}
\caption{Two-dimensional example showing the distance map from the circular GTV and the 1\% diffusion level (white lines) on top of each other.\label{FIG_circle}}
\end{figure}

Next a simple two-dimensional model is introduced: a circular GTV with a diameter of 2~cm. For this and the following {\cb 2D} examples, the grid resolution is set to $\Delta x = \Delta y = 0.1$~mm. The time step $\Delta t$ is chosen so that condition (\ref{EQ_timestep}) is fulfilled. Figure \ref{FIG_circle} shows the distance map calculated with the fast marching method. On top of it, it shows the 1\% iso-cell-density lines from the diffusion model at different time points (white lines). Unsurprisingly, both the fast marching and diffusion methods lead to uniform circular expansions in this isotropic geometry. As in the one-dimensional example, the speed of the wave front is approximately 1~cm per 100 days.  

\begin{figure}[ht]
\centering
\includegraphics[width=0.7\textwidth]{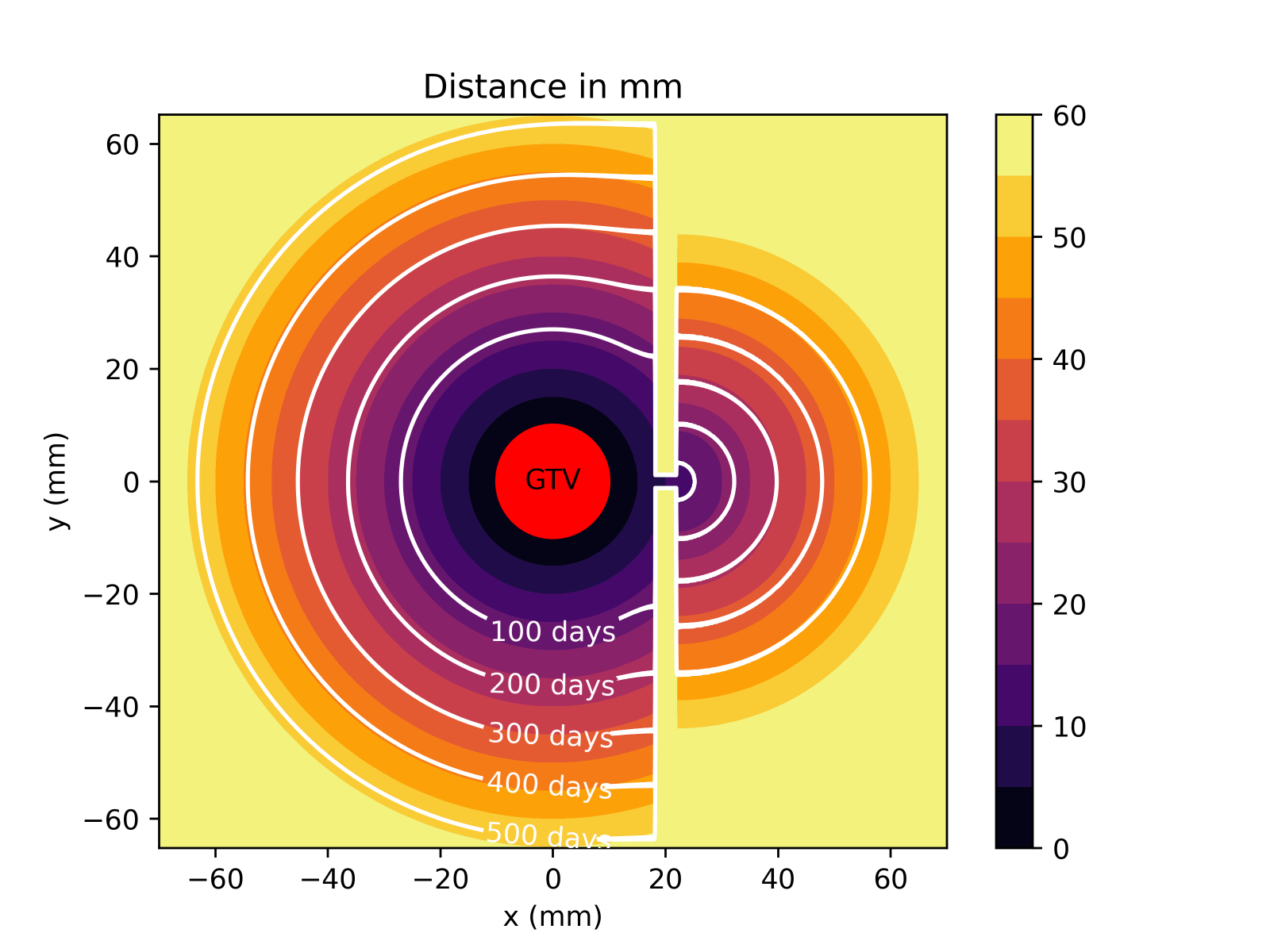}
\caption{Two-dimensional example with an obstacle (wall at $x=20$~mm) that has a small hole. The 1\% diffusion levels (white lines) are plotted on top of the distance map. \label{FIG_obstacle}}
\end{figure}

Let us now investigate how the shortest path and diffusion models behave in the presence of an obstacle with a ``pinhole''. This geometry represents an impermeable barrier to tumor spread with a small hole. The obstacle is modeled by setting the resistance to a very high value of $g=10^{14}$ within the obstacle. Correspondingly, the diffusion coefficient in the obstacle is reduced to $d = 0.25 \cdot 10^{-14} \, \text{mm}^2/\text{day}$. Figure \ref{FIG_obstacle} shows the 1\% tumor front on top of the distance map, similar to figure \ref{FIG_circle}. 

While the overall shapes are similar, two differences can be observed: first, the diffusion fronts (white lines) enter the obstacle always at a right angle. On the contrary, the orientation of the iso-distance bands is not affected near the obstacle, as seen on the left side of the wall. 

The second difference is that, unlike the diffusion fronts, the distance map is (by definition) not affected by the presence of the wall along the line at $y=0$ that goes through the hole. Whereas the iso-distance bands reach the same extent in the $x$-direction on both side of the wall  ($x \simeq  \pm 60$~mm), the diffusion fronts are perturbed on the right-side of the wall, resulting in an asymmetrical extent along the x-axis. The reason is that the wall reduces the diffusive flow to the right side, which in turn moves the 1\% cell density line inwards. 

For the final example, a more complex geometry with anisotropy in the tumor spread is introduced. The anisotropy can come from directional tissues that serve as highways for tumor cells, such as white matter tracts for brain tumors (corpus callosum) \citep{jordan2019},  and muscle fibers for certain types of sarcoma \citep{salerno2021}. 

\begin{figure}[ht]
\centering

\begin{subfigure}{0.6\textwidth}
\centering
\includegraphics[width=\textwidth]{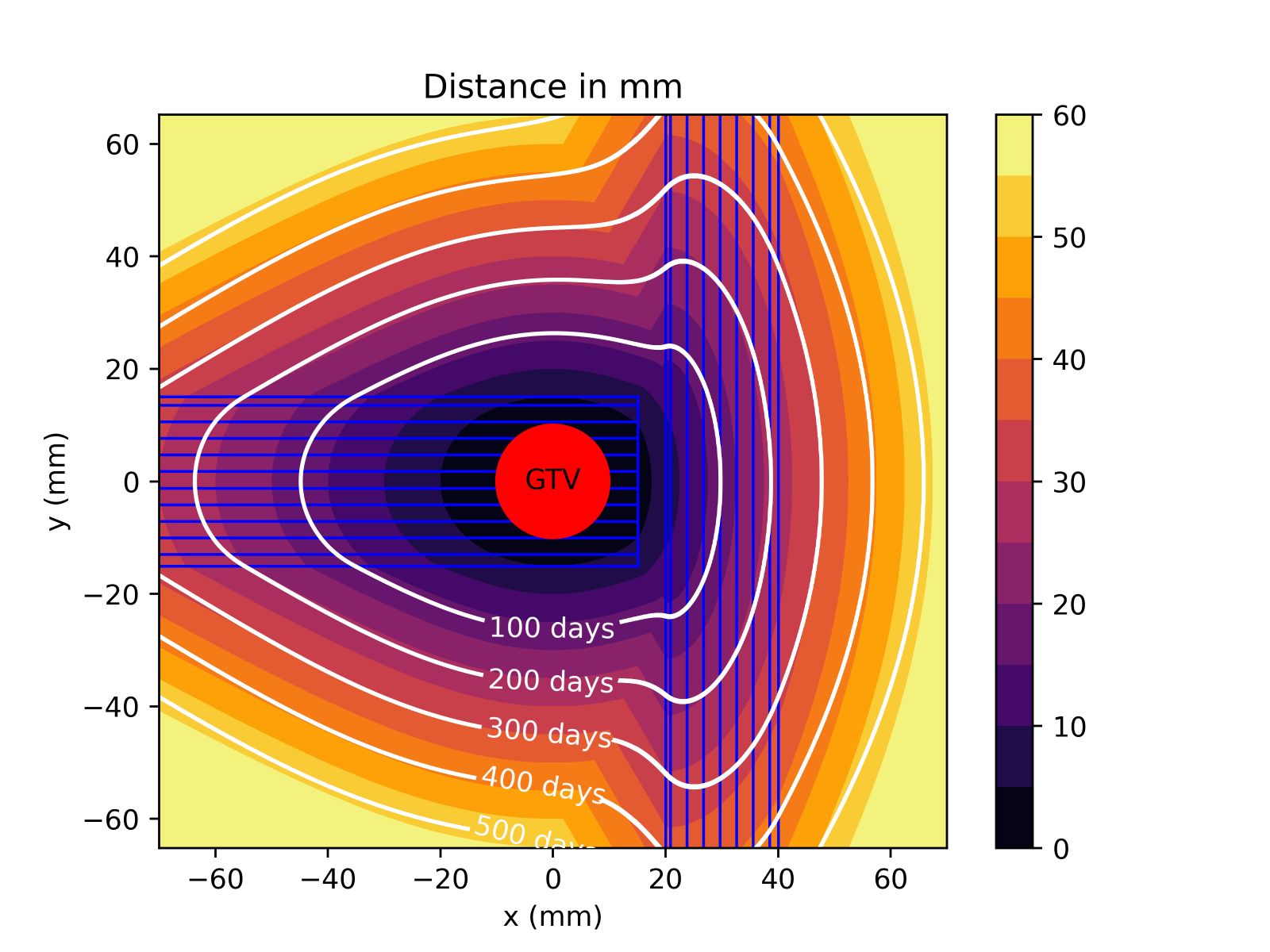}
\caption{}
\end{subfigure}

\begin{subfigure}{0.6\textwidth}
\centering
\includegraphics[width=\textwidth]{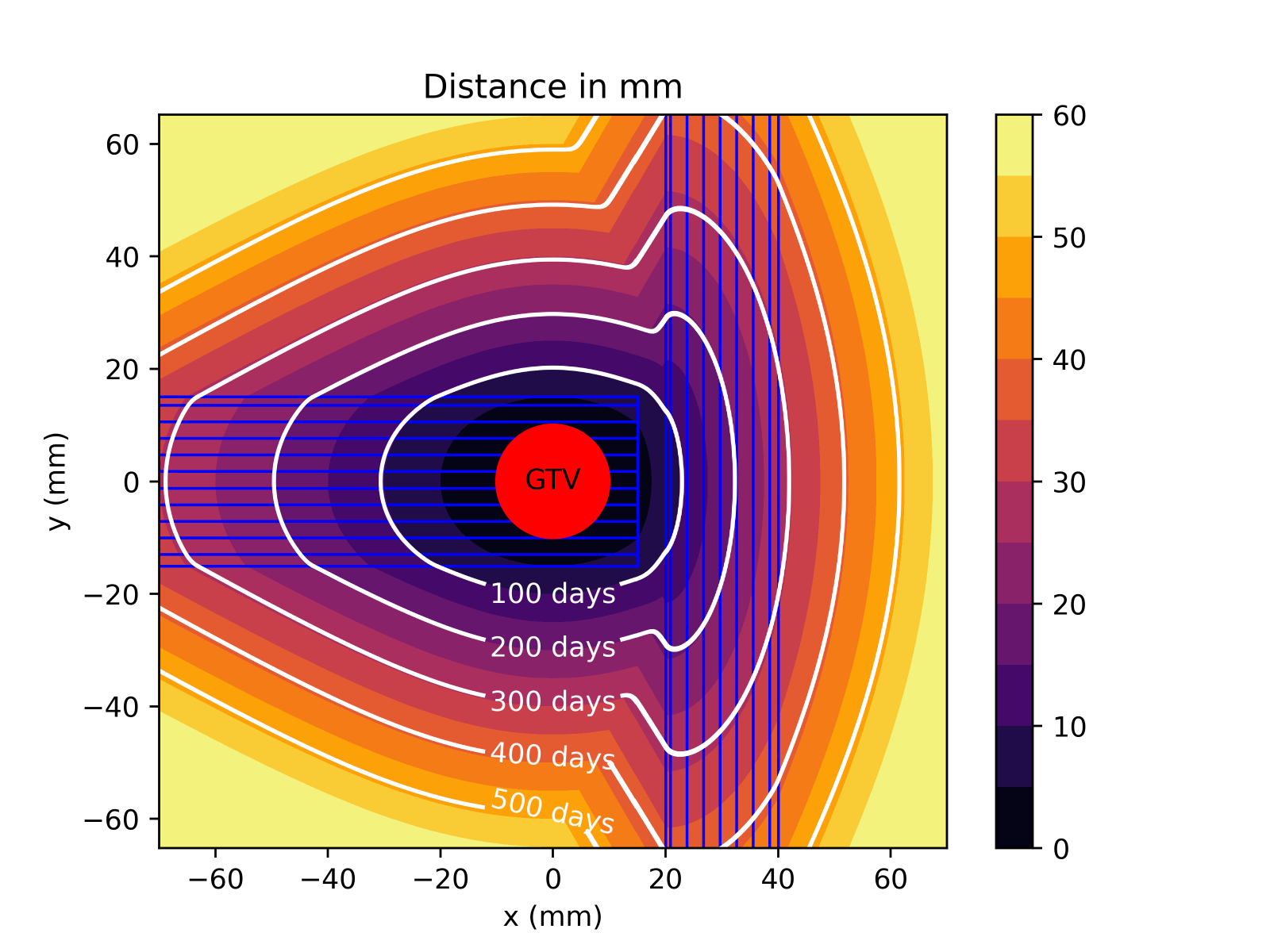}
\caption{}
\end{subfigure}

\caption{Geometric example with two structures exhibiting preferred tumor spread in the horizontal and vertical directions, shown by the blue lines. As in the previous examples, the 1\% cell density levels from the diffusion model (white lines) are plotted on top of the distance map. The $q$ parameter was set to the standard value of $q=5$~mm in (a) and to $q=0.5$~mm in (b). \label{FIG_aniso}}
\end{figure}

Figure \ref{FIG_aniso} shows two separate structures with preferred spread in the horizontal direction (left structure) and vertical direction (right structure). The orientation of the ``fibers'' is marked by the blue lines. Along those lines the resistance $g$ is reduced by a factor of 4, resulting in an expected two-fold expansion of the iso-distance bands. For example, in the structure with vertical orientation between $x=20$~mm and $x=40$~mm, the metric tensor is $ \mathcal{G} = \begin{pmatrix} 1 & 0\\0 & 0.25 \end{pmatrix}$. Correspondingly, the diffusion coefficient is {\em increased} by a factor of 4 in the direction of preferred spread. The diffusion tensor in that structure is $\mathcal{D} =    \begin{pmatrix} 0.25 & 0\\0 & 1 \end{pmatrix} \text{mm}^2/\text{day}$.

\begin{figure}[ht]
\centering
\includegraphics[width=0.7\textwidth]{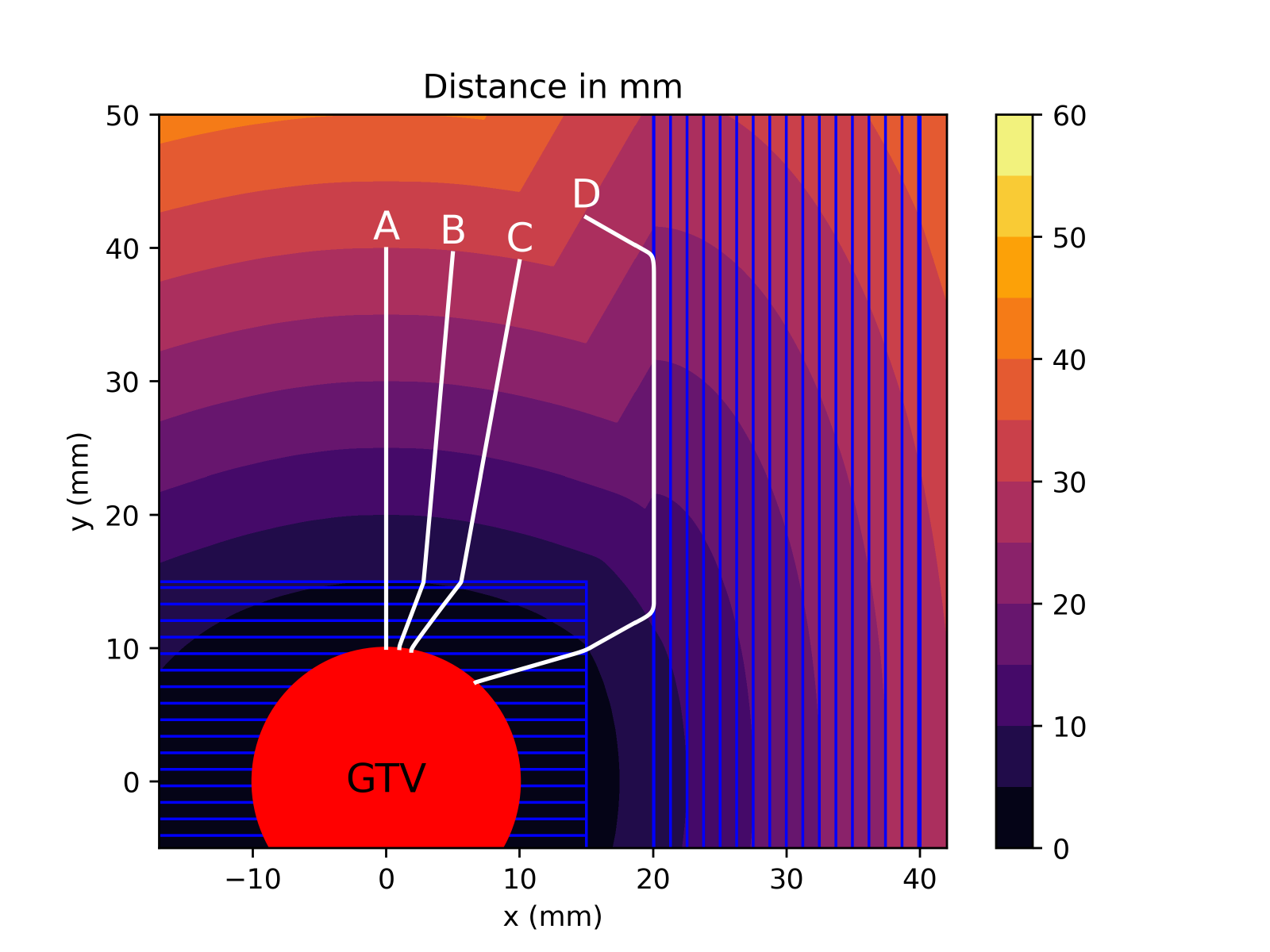}
\caption{This is a zoomed-in version of figure \ref{FIG_aniso} with the diffusion fronts removed. It also shows the shortest paths (geodesics) from the GTV to four points A-D. In homogeneous isotropic tissues, the geodesics are perpendicular to the iso-distance fronts (level sets of $S(\boldsymbol{r})$), while in anisotropic tissues and near boundaries they are not.  \label{FIG_geodesics}}
\end{figure}

As in the previous example, the overall  shape of the diffusion fronts follows the iso-distance bands. However, one obvious difference is that the distance bands show sharp kinks and corners, wheres the diffusion fronts are smooth with the standard $q$ parameter of $q=5$~mm in figure \ref{FIG_aniso}(a).  This finding can be explained with the help of figure \ref{FIG_geodesics}. If we look at the shortest paths (geodesics) from the GTV to four points that are all at the same distance $\sigma = 30$~mm but 5 mm apart from each other on the $x$ axis, we see an abrupt change in the shortest path between points A-C, and point D. The shortest path to point D follows a ``detour''  along the vertical fiber highway. This difference in the shortest path characteristics leads to the kink of the iso-distance band between point C and D. The diffusion process, on the other hand, leads to an effective averaging of the paths and therefore to smoother fronts. After reducing the  $q$ parameter from 5~mm to 0.5~mm, the diffusion fronts follow the distance bands much more closely, as shown in figure \ref{FIG_aniso}(b). 

\begin{figure}[ht]
\centering
\begin{subfigure}{0.48\textwidth}
\centering
\includegraphics[width=\textwidth]{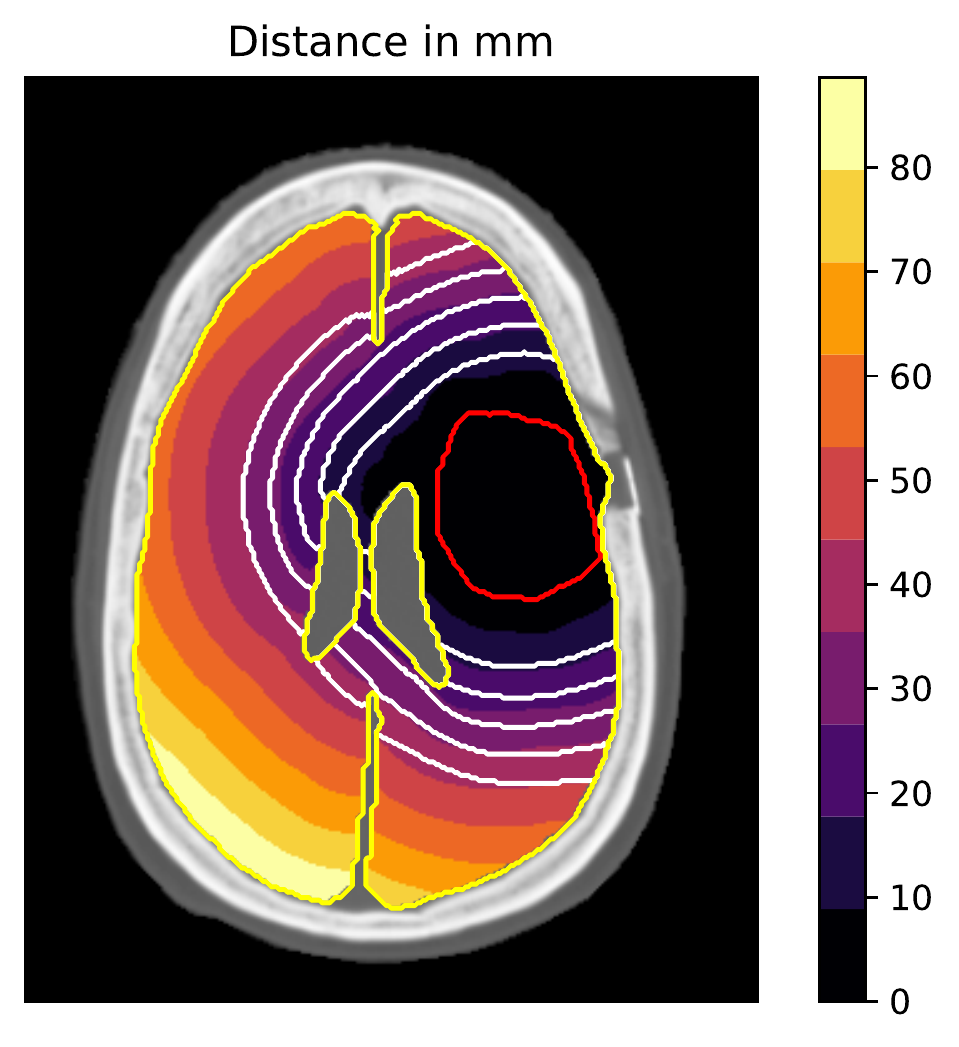}
\caption{}
\end{subfigure}

\begin{subfigure}{0.48\textwidth}
\centering
\includegraphics[width=\textwidth]{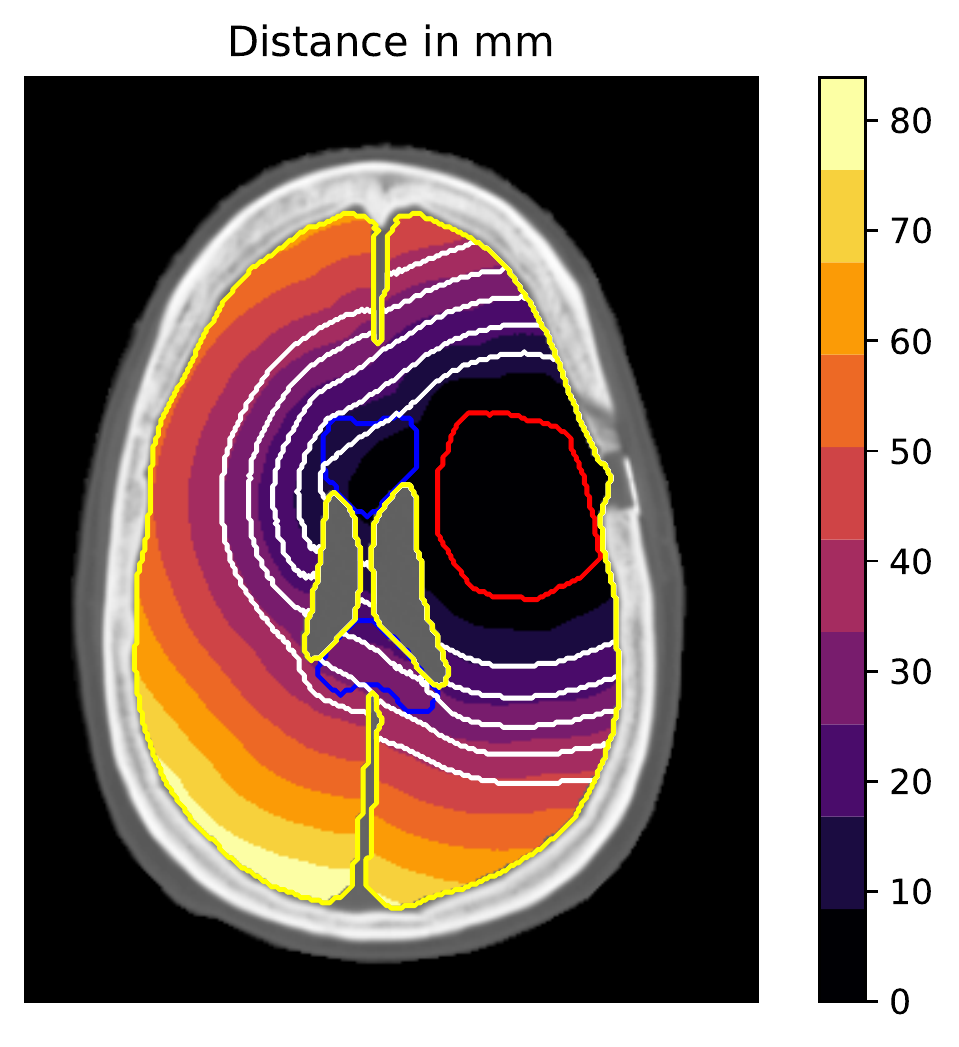}
\caption{}
\end{subfigure}

\caption{\cb 3D example of a GBM patient with barrier structures (falx cerebri, ventricles and skull). The distance map overlays the CT image. The 1\% diffusion levels (white lines) are plotted at different time points from 100 days (innermost contour) to 500 days (outermost contour) with a 100 day interval. The GTV and barrier structure delineations are displayed in red and yellow, respectively. Isotropic diffusion is spread is shown in (a) while a preferred path of spread in the corpus callosum (in blue) is simulated in (b). \label{FIG_3D_obstacle}}
\end{figure}

{\cb
Finally, the shortest path and diffusion models are applied to a realistic geometry of a glioblastoma multiforme (GBM) patient with barrier structures. The barrier structures for GBM include the ventricles, falx cerebri, tentorium cerebelli, brainstem, and the inner surface of the skull~\citep{niyazi2016}. In Fig. \ref{FIG_3D_obstacle}, the results are shown for isotropic diffusion in (a) and anisotropic diffusion in (b). As before, for the case of anisotropic diffusion, the diffusion coefficient in the corpus callosum is increased with a factor 4 along the $x$-direction, corresponding to the directionality of the nerve fibers. The same behaviour of the models is observed as in the 2D case by comparing the diffusion fronts with iso-distance bands, although the differences between the models become less pronounced. This can be explained by the fact that for realistic geometries, the edges of the structures are less sharp as compared to the idealized 2D examples above.
}

\section{Discussion}
The results show a high degree of similarity between two seemingly disparate models to determine tumor fronts beyond the visible GTV, namely the geometric expansion model and the  diffusion model.
The primary difference is that the diffusion model generally yields smoother contours. The degree of smoothness 
can be controlled by adjusting the infiltration length $q=\sqrt{d / \rho}$.  
By setting $q$  to a much smaller value such as $q=0.5$~mm or less, the diffusion fronts can be made to follow the  iso-distance bands of the geometric expansion model much more closely, to within arbitrary precision, even in complex anisotropic geometries such as the one shown in  figure \ref{FIG_aniso}. 

In terms of practical impact of these findings, one can foresee that a more mechanistically based model such as the diffusion model with the $q$ parameter to  control its smoothness could be attractive for GTV-CTV expansion in treatment planning. 
Another practical application could be for quality assurance of GTV-CTV expansion modules. Since the two completely independent methods with different algorithmic implementations such as fast marching and FTCS PDE (partial differential equation) solver must yield identical results in the limit $q \to 0$, one method can be used as an independent  check of the other.

{\cb The problem of estimating and personalizing the parameters of the diffusion model, in particular the proliferation rates $\rho$ and diffusion coefficients $d$, has been investigated in previous studies such as \cite{konukoglu2009} and \cite{baldock2013}, but it is far from being solved. The use of additional imaging modalities such as diffusion tensor imaging could help with the personalization \citep{peeken2019}). The sensitivity of the tumor fronts to uncertainties in the estimates of the underlying parameters also requires further investigation.}

\section{Conclusion}
Both the geometric expansion using fast marching and the diffusion model using a partial differential equation solver can be used to expand the gross tumor volume to the clinical target volume. The diffusion model generally leads to smoother contours. The degree of smoothness can be controlled by a parameter that equals the square root of the ratio of the diffusion coefficient and the proliferation rate.

\section*{Acknowledgments}
{\cb We} wish to thank Drs. Bram L Gorissen and Ali Ajdari for their thoughtful comments that have improved the manuscript. Research reported in this publication was supported by the National Cancer Institute of the United States under grant number R01CA266275, and by the Therapy Imaging Program (TIP) funded by the Federal Share of program income earned by Massachusetts General Hospital on C06CA059267, Proton Therapy Research and Treatment Center. The content is solely the responsibility of the authors and does not necessarily represent the official views of the National Institutes of Health.

{\cb
\appendix
\section{}
Here we show how to generalize equation (\ref{EQ_Eik}) to the anisotropic case, resulting in equation (\ref{EQ_Riemann_comp}).
We focus on a point $\boldsymbol{r} = \boldsymbol{r}_0$ with anisotropic metric $\mathcal{G}(\boldsymbol{r}_0)$. Let us drop $(\boldsymbol{r}_0)$ from subsequent equations.   
We introduce a new basis with coordinates $(u, v, w)$ such that (locally at point $\boldsymbol{r}_0$)
the distance becomes the Euclidean distance 
\begin{equation}
(ds)^2 = (du)^2 + (dv)^2 + (dw)^2 \label{EQ_arc_u}.
\end{equation}
Let $\mathcal{J}$ be the Jacobian matrix of this transformation: 
\begin{equation}
\mathcal{J} = \begin{pmatrix} 
\frac{\partial u}{\partial x} & \frac{\partial u}{\partial y} & \frac{\partial u}{\partial z}\\
\frac{\partial v}{\partial x} & \frac{\partial v}{\partial y} & \frac{\partial v}{\partial z}\\
\frac{\partial w}{\partial x} & \frac{\partial w}{\partial y} & \frac{\partial w}{\partial z}
\end{pmatrix}, \label{EQ_Jacobian}
\end{equation}
such that 
$(du, dv, dw)' = \mathcal{J} (dx, dy, dz)'$. 
Comparing equation (\ref{EQ_aniso}) with (\ref{EQ_arc_u}) we see that 
 $\mathcal{G} = \mathcal{J}' \cdot \mathcal{J}$. 
 
We can now express the shortest distance map $S(x,y,z)$ in $(u,v,w)$ coordinates and introduce the corresponding gradient ${\boldsymbol{\tilde \nabla}} 
S= (\frac{\partial S}{\partial u}, \frac{\partial S}{\partial v}, \frac{\partial S}{\partial w})'$. 

In the $(u,v,w)$ space with the unit matrix as the metric tensor, the iso-distant fronts move along the gradients as in equation (\ref{EQ_Eik}) with $g=1$,  such that:  
\begin{equation}
{\boldsymbol{\tilde \nabla}}'S \cdot  {\boldsymbol{\tilde \nabla}}S =1. \label{EQ_Eik_trans}
\end{equation}

Now from equation (\ref{EQ_Jacobian}) we see that 
\begin{equation}
\mathcal{J}' \cdot {\boldsymbol{\tilde \nabla}}S =   {\boldsymbol{\nabla}}S
\end{equation}
and therefore
\begin{equation}
{\boldsymbol{\tilde \nabla}}S =  \mathcal{J}'^{-1} \cdot  {\boldsymbol{\nabla}}S.
\end{equation}
Inserting this into equation (\ref{EQ_Eik_trans}) yields:
\begin{equation}
{\boldsymbol{\nabla}}'S \cdot \mathcal{J}^{-1} \cdot \mathcal{J}'^{-1} \cdot {\boldsymbol{\nabla}}S =1.
\end{equation}
Because $\mathcal{J}^{-1} \cdot \mathcal{J}'^{-1} =  \mathcal{G}^{-1} $, and the derivation above holds for any point $\boldsymbol{r} = \boldsymbol{r}_0$, we obtain equation (\ref{EQ_Riemann_comp}).}


\newcommand{\newblock}{}

\bibliographystyle{harvard/dcu} 
\bibliography{diffusion} 

\end{document}